\documentclass[useAMS,usenatbib]{mn2e}
\newcommand{\FIG}[1]{}

\usepackage{amssymb}
\usepackage{amsmath}
\usepackage{subfigure}
\usepackage{textcomp}
\usepackage{graphicx}
\usepackage{csquotes}

\title{Do fragmentation and accretion affect the stellar Initial Mass Function?}

\author[R. Riaz,  D.R.G. Schleicher, S. Vanaverbeke and Ralf S. Klessen]{R. Riaz$^{1}$\thanks{E-mail: rriaz@astro-udec.cl} 
D.R.G. Schleicher$^{1}$\thanks{E-mail: dschleicher@astro-udec.cl} S. Vanaverbeke$^{2}$\thanks{E-mail: siegfriedvanaverbeke@gmail.com} Ralf S. Klessen$^{3,4}$\thanks{E-mail: klessen@uni-heidelberg.de} \\
$^{1}$Departamento de Astronom\'ia, Facultad Ciencias F\'isicas y Matem\'aticas, Universidad de Concepci\'on, Av. Esteban Iturra s/n Barrio \\
Universitario, Casilla $160$-C, Concepci\'on, Chile \\
$^{2}$Centre for mathematical Plasma-Astrophysics, Department of Mathematics, KU Leuven, Celestijnenlaan 200B, 3001 Heverlee, Belgium \\ 
$^{3}$Universit{\"a}t Heidelberg, Zentrum f{\"u}r Astronomie, Institut  f{\"u}r Theoretische Astrophysik, Albert-Ueberle-Str. 2,  69120 Heidelberg, Germany \\
$^{4}$Universit{\"a}t Heidelberg, Interdisziplin{\"a}res Zentrum f{\"u}r Wissenschaftliches Rechnen, Im Neuenheimer Feld 205,  69120 Heidelberg, Germany
}

\begin{document}

\date{Accepted - Received -}

\pagerange{\pageref{firstpage}--\pageref{lastpage}} \pubyear{2015}

\maketitle

\label{firstpage}


\begin{abstract}
While the stellar Initial Mass Function (IMF) appears to be close to universal within the Milky Way galaxy, it is strongly suspected to be different in the primordial Universe, where molecular hydrogen cooling is less efficient and the gas temperature can be higher by a factor of 30. In between these extreme cases, the gas temperature varies depending on the environment, metallicity and radiation background. In this paper we explore if changes of the gas temperature affect the IMF of the stars considering fragmentation and accretion. The fragmentation behavior depends mostly on the Jeans mass at the turning point in the equation of state where a transition occurs from an approximately isothermal to an adiabatic regime due dust opacities. The Jeans mass at this transition in the equation of state is always very similar, independent of the initial temperature, and therefore the initial mass of the fragments is very similar. Accretion on the other hand is strongly temperature dependent. We argue that the latter becomes the dominant process for star formation efficiencies above 5$-$7 \%, increasing the average mass of the stars. 

\end{abstract}


\section{Introduction}

Star formation started with the first stars, which were expected to form from a primordial gas, where cooling was predominantly regulated via molecular hydrogen $H_{2}$ \citep{Saslaw67, Palla83, Galli98, Glover08, Clark2011a, Clark2011b}. While $H_{2}$ cooling never decreases the gas temperature below $T$ $\sim$ 200 K, the contribution from HD, however, decreases the gas temperature to $T$ $\sim$ 50$-$100 K \citep{Ripamonti2007, Bovino2014}. This is an indication that the Jeans mass ($M_{\rm J}$) was enhanced compared to present-day clouds, which was originally thought to lead to  stellar masses of 10$-$1000 $M_\odot$ \citep{Abel02, Bromm04, Yoshida08, Latif13, Latif15}. Taking into account the process of disk fragmentation and possible early ejection from the disk, the final masses of protostars, however, were found to be 0.1$-$10 $M_\odot$ \citep{Greif11}. The formation of low-mass stars in the primordial regime has also been suggested in earlier studies \citep[e.g.][]{Nakamura2002, Palla83}. Already at metallicities $Z \sim$ 10$^{-5}$ $Z_\odot$~ \citet{Clark2008} have reported vigorous fragmentation due to efficient  dust cooling that leads to densely-packed clusters of low-mass stars in which the IMF of stars peaks below 1 $M_\odot$.

With the ejection of metals, the stellar mass decreases due to the increasing efficiency of cooling. In particular, metal line cooling predominantly proceeds via oxygen, carbon and nitrogen lines \citep{Bromm03}. This cooling mechanism becomes quite efficient for metal abundances of about $0.3~  \%$ of the solar value, and operates already at low densities, of the order 1$-$100~cm$^{-3}$ depending on metallicity  \citep{Omukai01, Omukai05, Glover07, Bovino14, Safranek14}.

An alternative cooling mechanism is provided by the injection of dust via AGB stars or supernovae, which provides a cooling mechanism that operates even at $\sim0.01\%$ of the dust abundance in the solar neighborhood, but kicks in at much higher densities ($>10^6$~cm$^{-3}$) than the metal line cooling  \citep{Schneider03, Schneider10}. While most of the extremely metal poor stars still have carbon and oxygen abundances that are consistent with metal line cooling  \citep{Frebel15}, the discovery of DSS J102915+172927 \citep{Caffau11a, Caffau11b, Schneider12} suggests that dust cooling must be relevant in some cases  \citep{Klessen12, Bovino16}. \citet{Clark2008} focusing on star formation in the early universe at very low metallicities have suggested a connection between the presence of angular momentum  and fragmentation in self gravitating disks. They have concluded that it leads to the formation of the first stellar clusters.  However, even in the presence of additional cooling mechanisms via metals and dust, the gas temperature will never decrease below the limit defined via the cosmic microwave background (CMB). The latter provides a minimum gas temperature that depends on cosmic redshift. 

Once the metallicity reaches the point where cooling becomes efficient, the characteristic gas temperature will be driven by the temperature evolution of the CMB, or alternatively by other radiation backgrounds, for instance in or in the vicinity of starburst galaxies, or in general in regions that are exposed to stronger radiation fields, including part of the ISM of our own Galaxy. While in the present-day Universe the IMF of stars appears rather universal \citep{Kroupa02, Chabrier03, Kroupa03, Chabrier14}, some evolution will be expected in the context of star formation at high redshift, as the minimum temperature provided by the CMB will be enhanced. Based on indirect considerations a top-heavy IMF at zero-metallicity has been suggested by \citet{Bromm04}. Also, \citet{Clark2011b} have investigated the collapse of high-redshift halos with metal free turbulent gas that remains unaffected by the prior star formation, suggesting on average relatively smaller masses compared to previous studies. However, in their simulations the accretion  is  still ongoing,  and  the  system  is  still  young. The masses they report are most likely  not  the  final  masses  of  the  protostars. Hence, their mass estimates may still be higher than in the present day interstellar medium (ISM).

The IMF is influenced by both the process of fragmentation and the protostellar accretion inside a collapsing gas cloud. The fragmentation itself strongly depends on the evolving Jeans mass $M_{\rm J}$ in the gas cloud \citep{Larson05}. The Jeans mass $M_{\rm J}$ depends on the critical density where the equation of state (EOS) is changing, typically from an approximately isothermal behavior at lower densities to an adiabatic behavior at high densities, once that the dust opacity becomes relevant and efficiently suppresess cooling.  \citet{Clark2009} employed a polytropic EOS based on models of \citet{Omukai2005} with an effective adiabatic index $\gamma$ = 1.06 for simulations of a primordial gas with $Z$ = 0, $Z$ = 10$^{-6}$~$Z_{\odot}$, and 10$^{-5}$~$Z_{\odot}$ to study fragmentation. Their findings suggest that even for a purely primordial gas fragmentation occurs at 10$^{-10}$ g cm$^{-3}$ $\leq$~$\rho$~$\leq$ 10$^{-8}$ g cm$^{-3}$. However, the resulting mass function is biased towards higher masses. Also, the gas with $Z$ = 10$^{-6}~$$Z_{\odot}$ exhibits less efficient fragmentation with expected stellar masses of several tens of solar masses. 

Our simulation scheme is similar to \citet{Clark2008}, as we also include the effects of rotation and follow the collapse up to an increase by  $\sim$ ten orders of magnitude in the gas density. \citet{Jappsen2005} suggested that the supersonic  turbulence    in  self-gravitating molecular  gas  can generate  a  complex  network  of  interacting filaments and that the  overall  density  distribution  could be  highly  inhomogeneous.  They pointed out that turbulent  compression  can sweep  up  gas  in  some parts  of  the  cloud,  but  other  regions  become  rarefied.  \citet{Peters12} have shown that the latter is strongly related to the equation of state, and requires $\gamma\lesssim1$. The fragmentation behavior of the cloud and its ability to form stars hence depend strongly on the EOS. We consider in our present work gas with transsonic turbulence corresponding to a Mach number $\mathcal{M}$ = 1.0, and still we observe strong filamentary structures forming within the gas and hosting most of the protostars emerging in collapsing gas.  \citet{Lee2018a} have demonstrated that the peak of the mass spectrum after fragmentation is determined by the point where a transition from isothermal evolution to an adiabatic evolution occurs due to the dust opacities. Also, \citet{Lee2019} have discussed  the  classical  point  of  view  in which  the  fragmentation  of  a  medium  is  characterized  by  the  Jeans  instability. Even in the presence of magnetic fields, they found this to lead to a universal IMF. 


On the other hand, protostellar accretion is another mechanism in the star forming gas that may influence the IMF. The protostellar accretion rate in the self-gravitating regime can be estimated as the Jeans mass divided by the free-fall time, leading to $\dot M$ $\sim$ $M_{\rm Jeans}$/$t_{\rm ff}$ $\sim$ $c_{\rm s}$$^{3}$ $\sim$ $T^{3/2}$. In this we aim to quantify the importance of accretion which in our results clearly becomes significant over time. We explore this possibility in the context of simplified toy models. For this purpose, we assume that metal and dust cooling is sufficiently efficient, so that the gas temperature is set by the temperature of the CMB, potentially including contributions from other radiation fields. We are only interested in investigating approximate trends, and will assume an initially isothermal evolution, with a transition to the adiabatic regime once the gas becomes optically thick to the dust grains. We present a set of collapse simulations exploring initial gas temperatures from 10 to 50 K. From the relation $T_{\rm r}$ = 2.725~(1 + $z$), our selected range of temperature approximately corresponds to the redshift $z$ = 2.7$-$17.3, if interpreted to correspond to the temperature of the CMB. In case of a truly zero metallicity the gas would be expected to remain isothermal for even longer and we would not expect to see a transition due to dust opacities, which we model here. But we note that even at high redshift, metal enrichment is rapid and the first generation of stars is expected to be short lived. Also, there is evidence for the presence of dust at significantly higher redshift environments \citep{Wilkins16, Valiante09, Maiolino04}. We also refer to other studies for explorations of the primordial collapse \citep[see][and other studies cited above]{Riaz18c}. 

Our main goal is to assess the evolution of the characteristic mass of the protostars as a function of minimum temperature. To cover reasonable statistics we perform our tests with two different initial seed values which provide additional outcomes to analyze for our selected range of the initial gas temperatures. Our method is presented in section 2, while the EOS is discussed in section 3. The main results are summarized in section 4 and the conclusion is given in section 5.

        \begin{figure*} \label{fig:2}
	\centering
	\includegraphics[angle=0,scale=0.55]{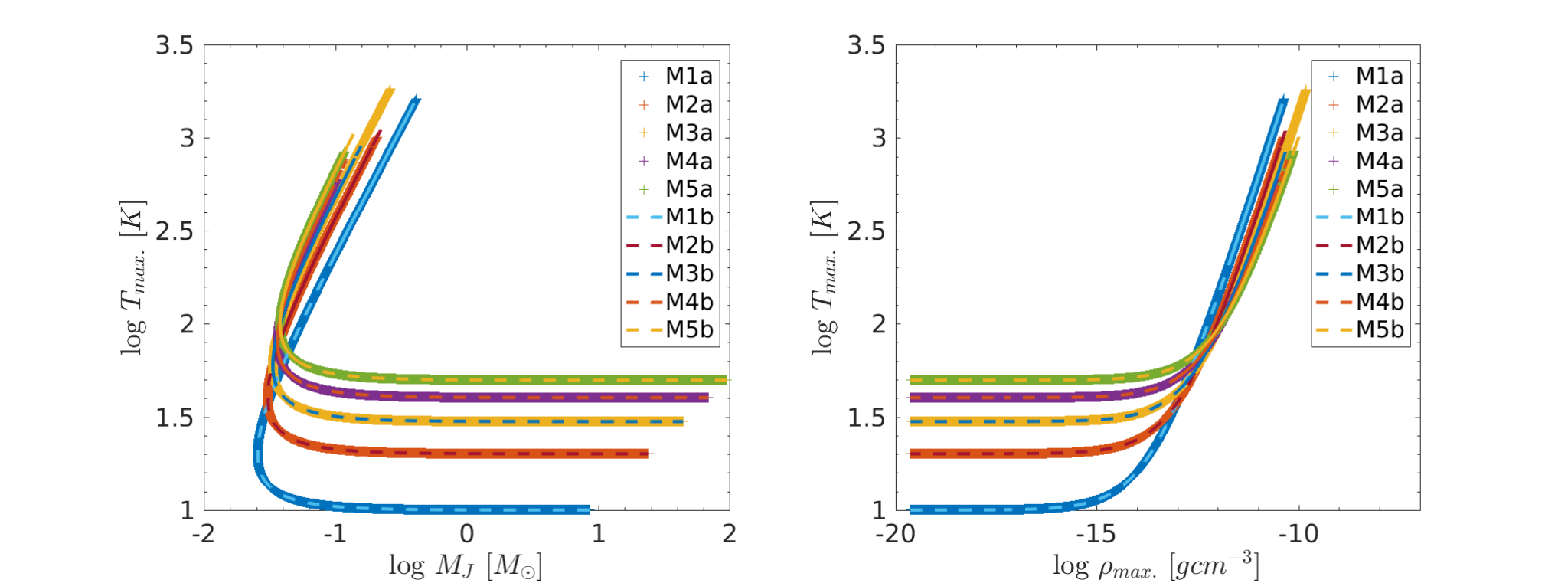}
	\caption{Left panel - evolution of temperature versus Jeans mass $M_{\rm J}$ of the collapsing cloud for models M1a$-$M5a and models M1b$-$M5b. Jeans mass $M_{\rm J}$ of the cloud is given in solar mass units and the temperature shown in Kelvin. Right panel - evolution of cloud temperature as a function of evolving density during the collapse of the cloud for models M1a$-$M5a and models M1b $-$M5b. Plus and dashed lines represent the results for first and second seed, respectively. Color in online edition.}
\end{figure*}

\begin{figure*} \label{fig:1}
    \centering
    \includegraphics[angle=0,scale=0.425]{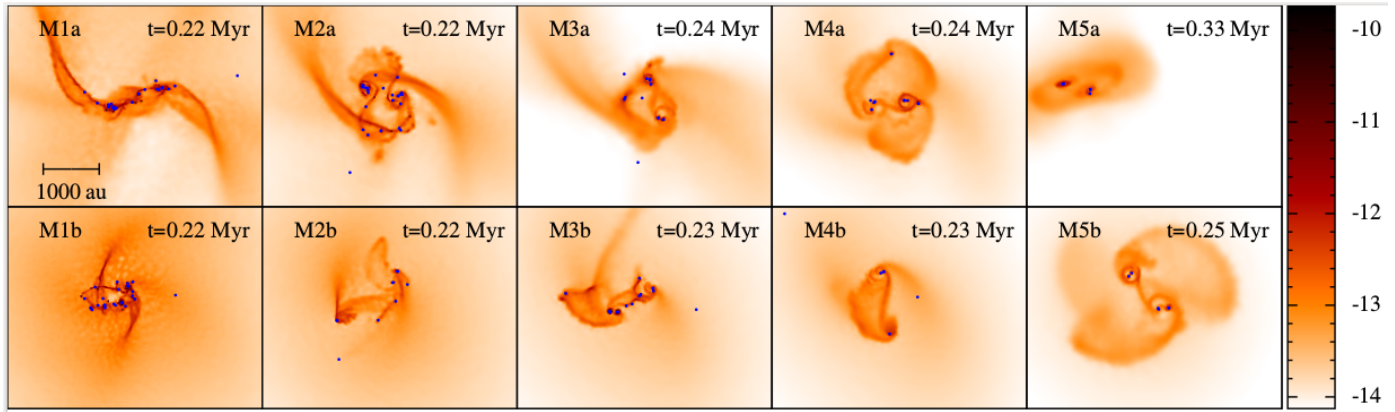}
    \caption{Simulation results for models M1a$-$M5a (top panels from left to right) and models M1b$-$M5b (bottom panels from left to right) at the end of the computed evolution of each model following the seed 1 and seed 2, respectively. Each panel shows a face-on view of the column density integrated along the z-axis. The shaded bar on the right shows log ($\rm \Sigma{}$) in g cm$^{-2}$. The corresponding dynamical time in Myr is shown at the top-right of each panel. Each calculation was performed with 250025 SPH particles. Also, these snapshots represent stage of evolution where star formation efficiency (SFE) in each model reaches $\xi$ = 15 \%. Color in online edition.}
  \end{figure*}

\section{Methodology}
The numerical models presented in this paper are based on smoothed particle hydrodynamics (SPH). We use the computer code GRADSPH \footnote{Webpage GRADSPH: http://www.swmath.org/software/1046} \citep{b46}. It is a fully three-dimensional SPH code which combines hydrodynamics with self-gravity and has been especially designed to study self-gravitating astrophysical systems such as molecular clouds \citep{b46}. The numerical scheme implemented in GRADSPH treats the long range gravitational interactions by using a tree-code gravity (TCG) scheme combined with the variable gravitational softening length method \citep{b33}, whereas the short range hydrodynamical interactions are solved using a variable smoothing length formalism. The code uses artificial viscosity to treat shock waves.

The evolution of the system of particles is computed using the second-order predict-evaluate-correct (PEC) scheme implemented in GRADSPH, which integrates the SPH form of the equations of hydrodynamics with individual time steps for each particle (see \citealt{b46}). GRADSPH has already been utilized successfully in recent studies related to star formation \citep{Riaz18a, Riaz18b, Riaz18c}. In our model setup the total mass inside the cloud is $M$ = 30 $M_{\odot}$ and the radius of the cloud is $R$ = 0.168 pc. The initial gas density is $\rho_{\rm i} = 1.0 \times 10^{-19}$~g~cm$^{-3}$. At that density, assuming a metallicity of at least 10$^{-1}$~$Z_{\odot}$, the gas may evolve in an approximately isothermal manner, as demonstrated e.g. by \citet{Omukai2005} and \citet{Grassi17}.       

For all models considered in the present work the gas is initially isothermal with a unique temperature assigned to each model. The free fall time is given as
\begin{equation} \label{freefalltime}
t_{\rm ff}=\sqrt{\frac{3 \pi}{32 G \rho_{\rm i}}}
\end{equation}
and is 0.188 Myr for the standard initial density $\rho_{\rm i}$. 
The initial condition is characterized by the parameters $\alpha$ and $\beta$, which correspond to the ratio of thermal and rotational energies with respect to the gravitational potential energy of the cloud. These parameters are defined as

\begin{equation} \label{alpha}
\alpha=\frac{5 R k T}{2 G M \mu m_{\rm H}},
\end{equation}

\begin{equation} \label{beta}
\beta=\frac{R^{3}\omega^{2}}{3 G M},
\end{equation} 
where $G$ is the gravitational constant, $k$ is the Boltzmann constant, $\mu = 2.33$ is the mean molecular weight, and $m_{\rm H}$ denotes the mass of the hydrogen atom. In our models, for the initial temperature ranging from 10 K to 50 K, the initial value of $\alpha$ varies between 0.115 and 0.578. The angular rotational frequency of the gas cloud is $\omega$ = 2.912 x 10$^{-14}$ rad s$^{-1}$. This sets the parameter $\beta{}$ = 1 \%~ which is kept fixed in all of our simulations. The initial setup is implemented in our SPH code by placing equal-mass particles on a hexagonal closely packed lattice and retaining only the particles within the initial cloud radius. The code uses internal dimensionless units which are defined by setting $G=M=R=1$.

We inject a spectrum of incompressible turbulence into the initial conditions in our models. We adopt two distinct values for the initial random seed while creating a velocity distribution for the injected turbulence to obtain two different realizations. This translates into two sets of models, M1a$-$M5a and M1b$-$M5b. The Mach number $\mathcal{M}$ is set to 1.0 so that the gas in each model is subject to transsonic turbulence. The dynamical time $t_{\rm dyn}$ of each model, which is equal to free fall time of the clouds is 0.188~ Myr. 
Our treatment of sink particles (protostars) employs recently introduced sink particle algorithm for SPH calculations described by \citet{b126}. We consider the merger of sink particles as described by \citet{b43}.
In our scheme, two sinks are allowed to merge if the following 3 criteria are met:
	
	\begin{itemize}\setlength{\itemsep}{0.25cm}
		\item Their relative distance $d$ is smaller then the accretion radius of the sinks (1 au) so that $d < r_{\rm acc}$.
		\item The total energy $E_{\rm tot}$ of the pair of sinks is negative, so that the pair is gravitationally bound.
		\item The least massive sink of the pair has insufficient angular momentum to remain rotationally supported against infall onto the primary i. e.  $j_{\rm sec}$ $<$ $j_{\rm cent}$, where $j_{\rm cent}$=$\sqrt{G~M_{\rm primary}~d}$ and $M_{\rm primary}$ denotes the mass of the most massive sink of the pair. 
	\end{itemize}

In the SPH framework the mass resolution is defined as $M_{\rm resolution}$ = 2 $N_{\rm opt}$ $m_{\rm particle}$. The number of SPH particles in each of our simulations is $N_{\rm SPH}$ = 250025 and for every SPH particle the number of neighboring particles is set as $N_{\rm opt}$ = 50. This indicates that our minimum resolvable mass is $M_{\rm resolvable} = 1.199 \times 10^{-2}$~$M_\odot$. We set a constant accretion radius $r_{\rm acc}$ = 1 au for the sink particles, and $r_{acc}$ always remains greater than the Jeans length in our simulations.

We aim to explore the effect of a change of the initial gas temperature assuming otherwise similar cloud properties. We focus particularly on trends regarding the IMF and explore the fragmentation and the mass accretion dominated regimes. To allow a consistent comparison of the models we terminate all of our simulations when a star formation efficiency (SFE) of $\xi$ = 15 \% is reached, where the SFE is defined as the ratio of protostellar mass to gas mass. We use the visualizing tool SPLASH \citep{SPLASH} which is publicly available to the community.

\begin{table*} \label{tbl-1}
\centering
\caption{Summary of the two sets of models M1a$-$M5a with the first random seed and M1b$-$M5b with the second random seed (corresponding to two different realizations of turbulence, with the same statistical properties). The entire table is constructed at the points in time when SFE $\xi$ reaches 2 \%, 5 \%, 10 \%, and 15 \% in each model. The table describes the ratio of kinetic to gravitational potential energy of the cloud ($\alpha$), the initial temperature ($T_{\rm i}$), the Jeans mass of the cloud evaluated at the initial gas density ($M_{\rm J}$), the total number of protostars produced ($N_{\rm max}$), the final number of protostars after mergers ($N_{\rm proto}$), the mean mass of protostars ($M^{\ast}_{\rm mean}$) with its variance, median mass of protostars ($M^{\ast}_{\rm median}$) with its variance, the lowest mass protostar ($M^{\ast}_{\rm min}$) and the highest mass protostar ($M^{\ast}_{\rm max}$). The initial radius, mass, and average density in each model are given as 0.168 pc, 30 $M_{\odot}$, and 10$^{-19}$ g cm$^{-3}$, respectively. The error estimation is performed with a confidence interval of 68.3 \%.  }

\begin{tabular}{cccccccccccc}
\multicolumn{10}{|c|}{$\xi$ = 2 \%}\\
\hline
\hline
Model & $\alpha$ & $T_{\rm i}$ (K) & $M_{\rm J}$ ($M_{\odot}$) & $N_{\rm max}$ & $N_{\rm proto}$  & $M^{\ast}_{\rm mean}$ ($M_{\odot}$) & $M^{\ast}_{\rm median}$ ($M_{\odot}$)  & $M^{\ast}_{\rm min}$ ($M_{\odot}$) & $M^{\ast}_{\rm max}$ ($M_{\odot}$) \\
\hline
M1a  & 0.115     & 10  &   4.855 & 15 & 5 & 0.120$\pm$ 0.036 & 0.117$\pm$ 0.036  & 0.036 & 0.252 \\
M2a  & 0.231     & 20  &   9.710 & 5 & 2 & 0.300$\pm$ 0.095 &  0.300$\pm$ 0.095  & 0.165  & 0.434 \\
M3a  & 0.347	 & 30  &  14.565 & 7 &  3 & 0.200$\pm$ 0.200 & 0.200$\pm$ 0.200 & 0.095  & 0.305 \\
M4a  & 0.463     & 40  &  19.420 & 8 & 7 & 0.086$\pm$ 0.031 &  0.078$\pm$ 0.031  & 0.0002  & 0.228 \\
M5a  & 0.578     & 50  &  24.275 & 10 &  5 & 0.120$\pm$ 0.042 &0.123$\pm$ 0.042  & 0.024  & 0.280 \\
M1b  & 0.115     & 10  &   4.855 & 16 & 10 & 0.060$\pm$ 0.018 &0.056$\pm$ 0.018 & 0.0002  & 0.113  \\
M2b  & 0.231     & 20  &   9.710 & 12 & 10& 0.060$\pm$ 0.020 & 0.046$\pm$ 0.020  & 0.0002 &  0.185 \\
M3b  & 0.347	 & 30  &  14.565 & 9 & 6 & 0.100$\pm$ 0.010 &  0.096$\pm$ 0.010 & 0.072  & 0.153 \\
M4b  & 0.463     & 40  &  19.420 & 18 & 8 & 0.075$\pm$ 0.018 & 0.058$\pm$ 0.018 & 0.027  & 0.175 \\
M5b  & 0.578     & 50  &  24.275 & 11 & 8  & 0.075$\pm$ 0.016 &0.076$\pm$ 0.016 & 0.017 & 0.170 \\
\hline
\multicolumn{10}{|c|}{$\xi$ = 5 \%}\\
\hline
\hline
M1a  & 0.115     & 10  &   4.855 & 15 & 5 & 0.300$\pm$ 0.104 & 0.303$\pm$ 0.104  & 0.029 & 0.683 \\
M2a  & 0.231     & 20  &   9.710 & 19 & 12 & 0.125$\pm$ 0.049 & 0.053$\pm$ 0.049  & 0.012  & 0.630 \\
M3a  & 0.347	 & 30  &  14.565 & 7 &  4 & 0.375$\pm$ 0.049 & 0.406$\pm$ 0.049 & 0.216  & 0.472  \\
M4a  & 0.463     & 40  &  19.420 & 10 & 5 & 0.300$\pm$ 0.026 & 0.324$\pm$ 0.026  & 0.190  & 0.354 \\
M5a  & 0.578     & 50  &  24.275 & 5 &  3 & 0.500$\pm$ 0.049 & 0.520$\pm$ 0.049  & 0.388  & 0.592 \\
M1b  & 0.115     & 10  &   4.855 & 16 & 8  & 0.187$\pm$ 0.029 & 0.183$\pm$ 0.029 & 0.063  & 0.309  \\
M2b  & 0.231     & 20  &   9.710 & 13 & 9 & 0.166$\pm$ 0.044 & 0.163$\pm$ 0.044  & 0.012  &  0.411 \\
M3b  & 0.347	 & 30  &  14.565 & 14 & 11 & 0.136$\pm$ 0.032 & 0.170$\pm$ 0.032 & 0.017  & 0.281 \\
M4b  & 0.463     & 40  &  19.420 & 18 & 5 & 0.300$\pm$ 0.050 & 0.240$\pm$ 0.050 & 0.186  & 0.463 \\
M5b  & 0.578     & 50  &  24.275 & 11 & 5  & 0.300$\pm$ 0.312 & 0.347$\pm$ 0.312 & 0.054 & 0.446 \\
\hline
\multicolumn{10}{|c|}{$\xi$ = 10 \%}\\
\hline
\hline
M1a  & 0.115     & 10  &   4.855 & 31 & 24 & 0.166$\pm$ 0.051 & 0.035$\pm$ 0.051  & 0.0002 & 0.856 \\
M2a  & 0.231     & 20  &   9.710 & 19 & 15 & 0.200$\pm$ 0.070 & 0.033$\pm$ 0.070  & 0.0002 & 0.865 \\
M3a  & 0.347	 & 30  &  14.565 & 17 &  9 & 0.333$\pm$ 0.088 & 0.270$\pm$ 0.088 & 0.0181  & 0.798  \\
M4a  & 0.463     & 40  &  19.420 & 12 & 6 & 0.500$\pm$ 0.087 & 0.611$\pm$ 0.087  & 0.1160  & 0.682 \\
M5a  & 0.578     & 50  &  24.275 & 12 &  9 & 0.460$\pm$ 0.133 & 0.316$\pm$ 0.133  & 0.0157  & 0.994 \\
M1b  & 0.115     & 10  &   4.855 & 19 & 17  & 0.177$\pm$ 0.029 & 0.077$\pm$ 0.029 & 0.0002  & 0.718  \\
M2b  & 0.231     & 20  &   9.710 & 20 & 16 & 0.189$\pm$ 0.044 & 0.085$\pm$ 0.044  & 0.0002  &  0.715 \\
M3b  & 0.347	 & 30  &  14.565 & 14 & 8 & 0.375$\pm$ 0.032 & 0.370$\pm$ 0.032 & 0.0347  & 0.943 \\
M4b  & 0.463     & 40  &  19.420 & 18 & 6 & 0.500$\pm$ 0.050 & 0.507$\pm$ 0.050 & 0.0258  & 1.269 \\
M5b  & 0.578     & 50  &  24.275 & 11 & 7  & 0.429$\pm$ 0.312 & 0.621$\pm$ 0.312 & 0.0038 & 0.741 \\
\hline
\multicolumn{10}{|c|}{$\xi$ = 15 \%}\\
\hline
\hline
M1a  & 0.115     & 10  &   4.855 & 26 & 23 & 0.157$\pm$ 0.0574 & 0.0474$\pm$ 0.0574  & 0.0181 & 1.0291 \\
M2a  & 0.231     & 20  &   9.710 & 31 & 30 & 0.1512$\pm$ 0.0437 & 0.0312$\pm$ 0.0437  & 0.0002  & 0.9262 \\
M3a  & 0.347	 & 30  &  14.565 & 17 & 12 & 0.3750$\pm$ 0.1130 & 0.1796$\pm$ 0.1130 & 0.0220  & 1.2935  \\
M4a  & 0.463     & 40  &  19.420 & 23 & 9 & 0.5000$\pm$ 0.1050 & 0.3648$\pm$ 0.1050  & 0.0594  & 0.9562 \\
M5a  & 0.578     & 50  &  24.275 & 12 &  9 & 0.5000$\pm$ 0.1430 & 0.376$\pm$ 0.1430  & 0.0157  & 1.0777 \\
M1b  & 0.115     & 10  &   4.855 & 28 & 24  & 0.1875$\pm$ 0.0420 & 0.0937$\pm$ 0.0420 & 0.0002  & 0.7116  \\
M2b  & 0.231     & 20  &   9.710 & 20 & 12 & 0.3750$\pm$ 0.1070 & 0.1796$\pm$ 0.1070  & 0.0197  &  0.9906 \\
M3b  & 0.347	 & 30  &  14.565 & 19 & 14 & 0.3214$\pm$ 0.1020 & 0.0704$\pm$ 0.1020 & 0.0121  & 1.2542 \\
M4b  & 0.463     & 40  &  19.420 & 31 & 15 & 0.3000$\pm$ 0.1060 & 0.1176$\pm$ 0.1060 & 0.0260  & 1.5367 \\
M5b  & 0.578     & 50  &  24.275 & 11 & 7  & 0.6429$\pm$ 0.1430 & 0.8763$\pm$ 0.1430 & 0.0078 & 1.0579 \\
\hline
\end{tabular}
\end{table*}

   \begin{figure*} \label{fig:3}
    \centering
    \includegraphics[angle=0,scale=0.45]{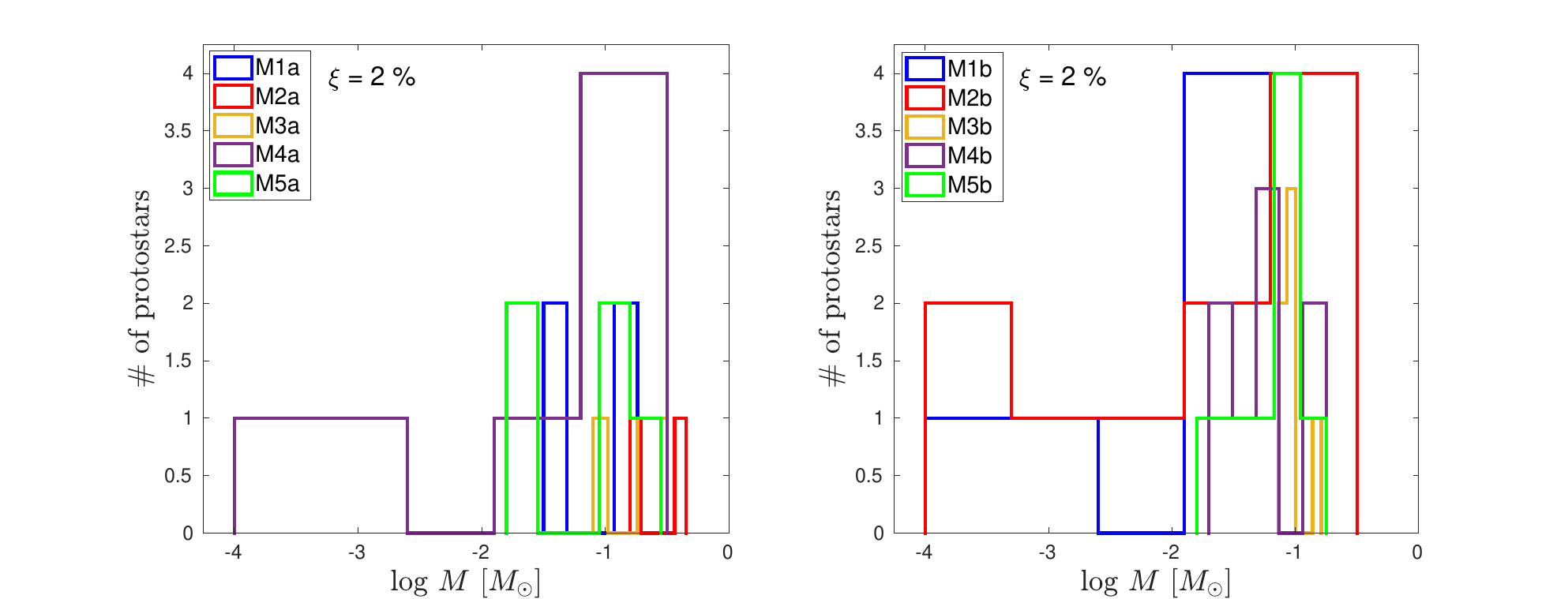}
    \includegraphics[angle=0,scale=0.45]{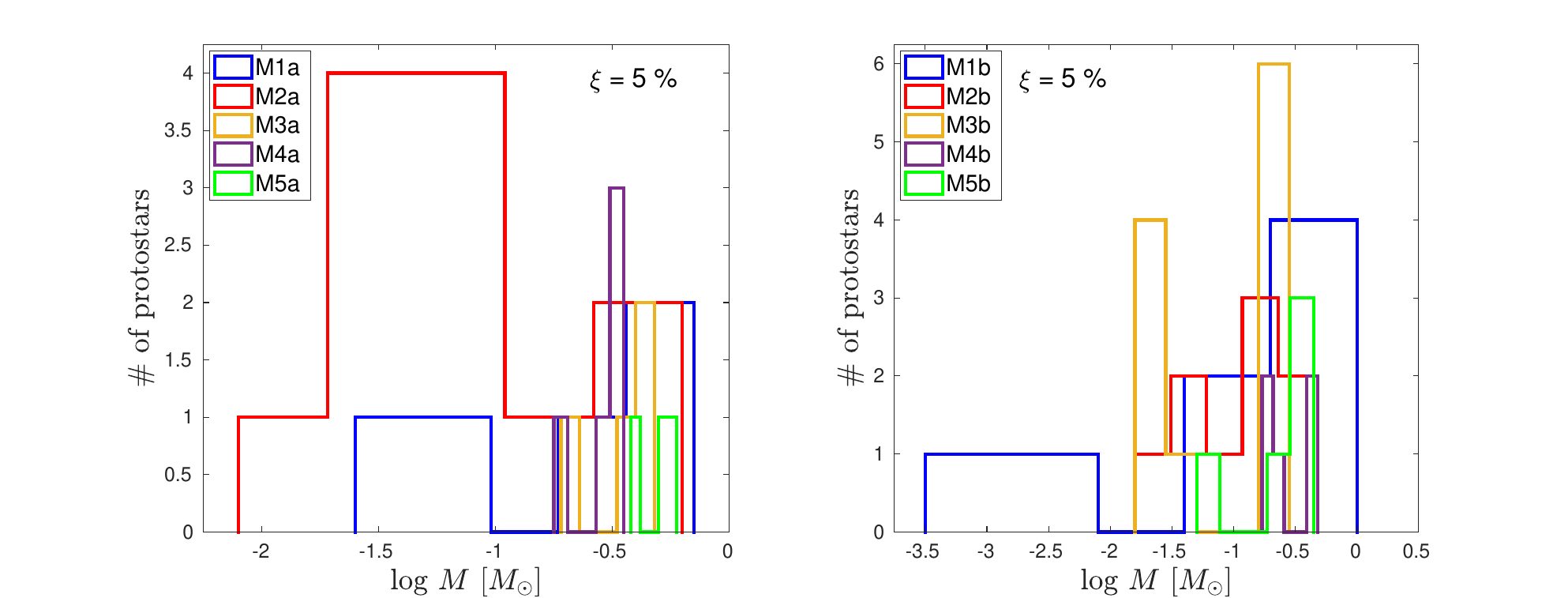}
    \includegraphics[angle=0,scale=0.45]{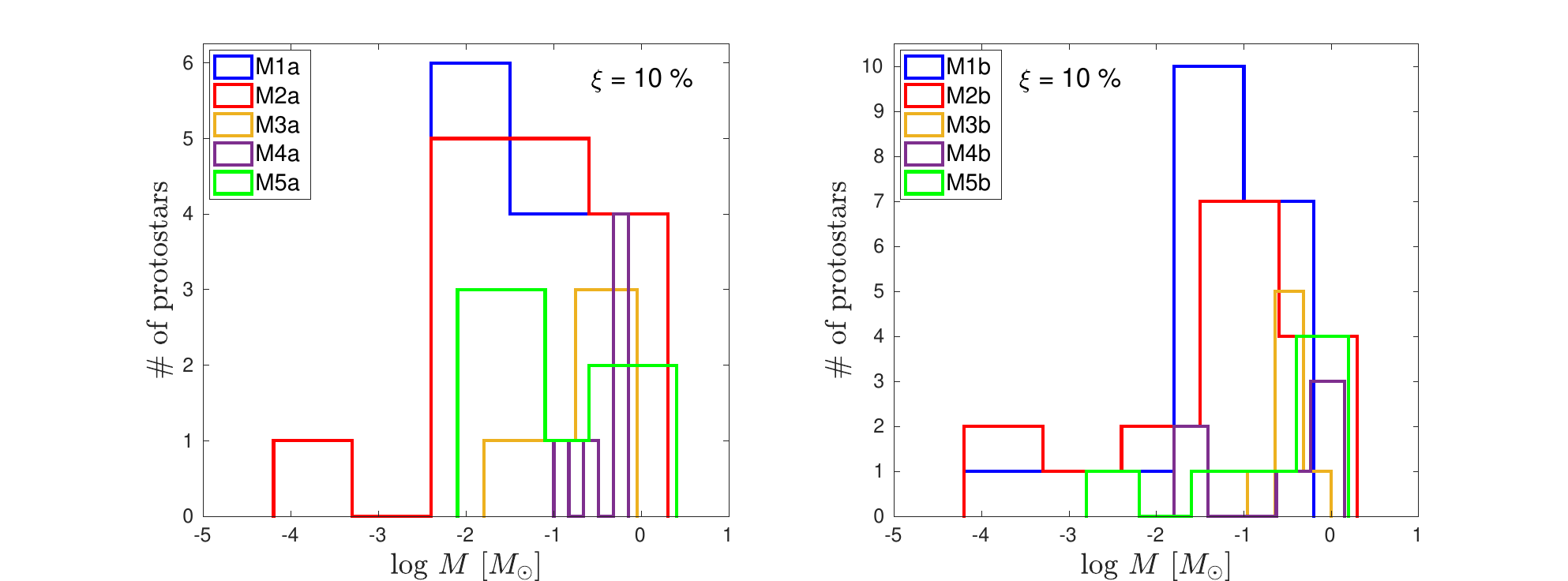}
    \includegraphics[angle=0,scale=0.45]{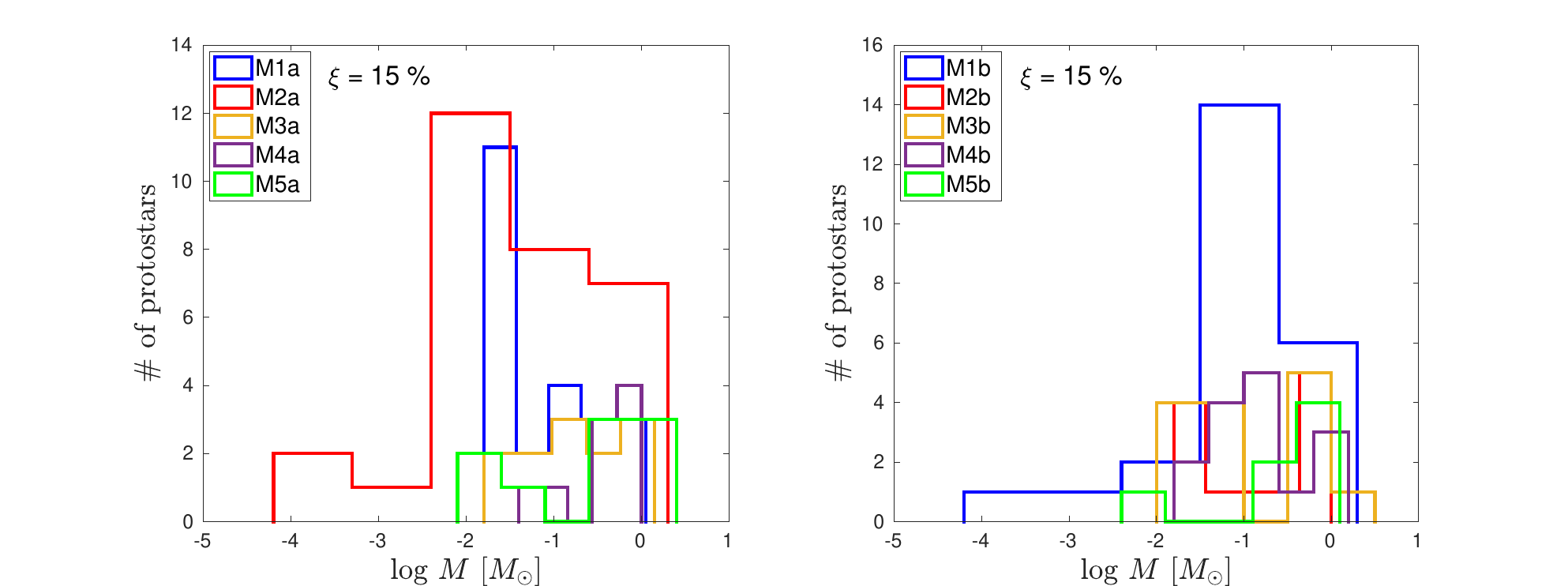}
    \caption{Mass distribution of the protostars formed during the collapse of models M1a$-$M5a (left-panels) and models M1b$-$M5b (right-panels). From the first to fourth row the mass distribution is presented at  points in time when the SFE $\xi$ reaches 2 \%, 5 \%, 10 \%, and 15 \%, respectively in each model. The masses are indicated in solar mass units. Color in online edition.}
  \end{figure*}
  



\section{Equation of state} \label{EOs}

We adopt a barotropic equation of state of the form
\begin{equation} \label{EOS}
P=\rho c_{0}^{2}\left[1+\left(\frac{\rho}{\rho_{\rm crit}}\right)^{\gamma-1}\right].
\end{equation}
To derive the values for the critical density \textit{$\rho$}$_{\rm crit}$ at which the phase transition from isothermal to adiabatic takes place in each of our models we use equation 20 of \citet{Omukai2005}. We consider the balance between the cooling which is dominated by continuum emission via dust thermal emission and compressional heating, implying  

\begin{equation} \label{transition}
T=\left(\frac{k^{3}}{12\sigma^{2}m_{\rm H}}\right) ^{1/5} {n^{2/5}_{\rm H}},
\end{equation}
where $k$, $\sigma$, $m_{\rm H}$, and $n_{\rm H}$ are the Boltzmann constant, Stefan-Boltzmann constant, mass of the hydrogen atom, and the number density of the gas, respectively. For the mass density, we have
\begin{equation} \label{transition}
\rho=n\mu m_{\rm H},
\end{equation}
which leads to the following expression to determine the critical density \textit{$\rho$}$_{crit}$:
\begin{equation} \label{transition}
\rho_{\rm crit}=6.115\times10^{-16}~T^{5/2}.   
\end{equation}
For each set of models we consider  $T_{\rm init}$ of 10 K, 20 K, 30 K, 40 K, and 50 K corresponding to critical densities \textit{$\rho$}$_{\rm crit}$ of $1.933 \times 10^{-13}$ g cm$^{-3}$, $1.094\times 10^{-12}$ g cm$^{-3}$, $3.013 \times 10^{-12}$ g cm$^{-3}$, $6.188\times 10^{-12}$ g cm$^{-3}$, and $1.080 \times 10^{-11}$ g cm$^{-3}$, respectively. The Jeans mass $M_{\rm J}$ \citep{Jeans1929} is defined as
\begin{equation} \label{transition}
M_{\rm J}=\frac{\pi}{6} \left(\frac{k}{\mu m_{\rm H}G} \right) ^{3/2} \rho^{-1/2}~ T^{3/2} ~ \sim \rho^{-1/2}~ T^{3/2}.
\end{equation}
The density evolves during the gas collapse until it reaches the critical value \textit{$\rho$}$_{\rm crit}$, and we get
\begin{equation} \label{transition}
M_{\rm J, crit}=(constant)~ T^{1/4}.
\end{equation}
From this result, we can see that at best a very weak dependence on the gas temperature can be expected in the fragmentation-dominated regime, where a factor of 10 in the gas temperature changes the Jeans mass by less than a factor of 2.

In Figure 1 the left and the right panels show the evolution of the thermal state of the cloud as a function of its evolving Jeans mass $M_{\rm J}$ and the evolution of the gas temperature as a function of evolving gas density in the two sets  of models M1a$-$M5a and M1b$-$M5b. The solid and the dashed lines represent results from the first and second seed, respectively. On the left, during the initial isothermal phase of the collapse the Jeans mass during the evolution of all models remains a decreasing function of gas density. This allows more clumps of gas to form and to become self-gravitating objects. However, after the change of the equation of state  at the critical density defined in equation 7 the evolution  leads to an increase of the Jeans mass $M_{\rm J}$ when the EOS is adiabatic. The smallest possible value of the Jeans mass $M_{\rm J,min}$ during the isothermal phase of the collapse shows a very weak dependence on the initial thermal state of the prestellar cloud as for higher initial temperatures, the transition occurs at a higher density, thereby largely compensating for the temperature effect. In our five models, we find a very small trend for the minimum thermal Jeans mass $M_{\rm J,min}$. For the lowest thermal state of $T_{\rm init}$ = 10 K set in models M1a and M1b we find that the Jeans mass $M_{\rm J}$ can attain a minimum value of 0.025 $M_{\odot}$, whereas for the warmest cloud $T_{\rm init}$ = 50 K explored in models M5a and M5b the smallest possible value of  $M_{\rm J,min}$ is 0.038 $M_{\odot}$. 

We note that the Jeans mass $M_{\rm J}$ at the transition density $\rho_{\rm crit}$ is always very similar, independent of the previous evolution. While a very weak dependence can be expected based on these considerations, a still much larger sample leading to better statistics would be needed to show such a dependence in practice. 
 
\section{Results and Discussion}

We present a summary of the evolution of all models in Table 1. As the simulations proceed we keep track of the protostellar mass evolution with respect to the evolving SFE in each of the gas clouds. We focus on the protostars that emerge inside the collapsing gas and also register the number of protostars that survive until the final stage of evolution when the SFE becomes $\xi$ = 2 \%, 5 \%, 10 \% and 15 \%. We notice that at $\xi$ = 2 \%, no correlation with the gas temperature can be recognized. $\xi$ = 5 \% is a borderline case where a weak trend might be visible, while at $\xi$ = 10 \% and 15 \%, it is clear that the larger average masses occur in the warmer models. 

In Figure 2 we show a sequence of the face-on views describing the final states of models M1a$-$M5a (top) and of the models M1b$-$M5b (bottom). The final details of each set of model outcomes along with the respective initial conditions are described in Table 1, which shows the results for the SFE $\xi$ = 15 \%. Table 1 provides information about the mass distribution at the intermediate stage, corresponding to $\xi$ = 2 \%, 5 \%, 10 \%, and 15 \%, respectively.

 Considering the top and bottom sequences of the panels in Figure 2, it seems evident that the gas temperature regulates the molecular gas that collapses under self-gravity. Relatively warmer gas results in more massive sinks but lesser fragmentation and the overall structure is more filamentary for colder gas. This feature remains visible in Figure 2 (both top and bottom) where the gas in the last panel  of each sequence corresponding to an initial temperature of 50 K yields a relatively small number of protostars i.e., $N_{\rm max}$ = 12 and 11, respectively, with mean masses of $M^{\ast}_{\rm mean}$ = 0.500$\pm$ 0.143 $M_{\odot}$ and 0.643$\pm$ 0.143 $M_{\odot}$. The number of protostars due to dynamical interactions and merger events becomes $N_{\rm proto}$ = 9 and 7, respectively. Contrary to this, the coldest of the clouds with an initial gas temperature of 10 K shown in the first panel of each sequence produce a total number of  $N_{\rm max}$ = 26 and 28 protostars, respectively, with respective mean masses $M^{\ast}_{\rm mean}$ = 0.157$\pm$ 0.057 $M_{\odot}$ and 0.187$\pm$ 0.042 $M_{\odot}$. In this particular set of models the number of protostars due to the dynamical interactions and merger events becomes $N_{\rm proto}$ = 23 and 24, respectively. 
 
  Filamentary structures are the formation sites of embedded protostars \citep[e.g.][]{Andre10}. We observe strong and well resolved filamentary gas structures inside cold gas clouds which remain under the influence of self-gravity. However, shifting from cold to warm gas clouds we find that the filamentary structures appear less resolved and are more blurred. Our results suggest that warmer clouds with $T_{\rm i}$ $\geq$ 30 K produce more massive protostars for sufficiently high SFEs (see Table 1 for $\xi$ = 15 \%).   

In Figure 3 we show for the two random seeds of the turbulence the resulting mass distributions of the protostars appearing in all five models when the SFE reaches $\xi$ = 2 \%, 5 \%, 10 \%, and 15 \%, respectively. In these figures, it becomes more visible at SFEs of 10 \%~ and 15 \% that the number of protostars is higher for colder gas, and that the fragment masses tend to be smaller in these cases, while no clear trends are recognized for lower SFEs.

            \begin{figure*} \label{fig:10}
    \centering
    \includegraphics[angle=0,scale=0.515]{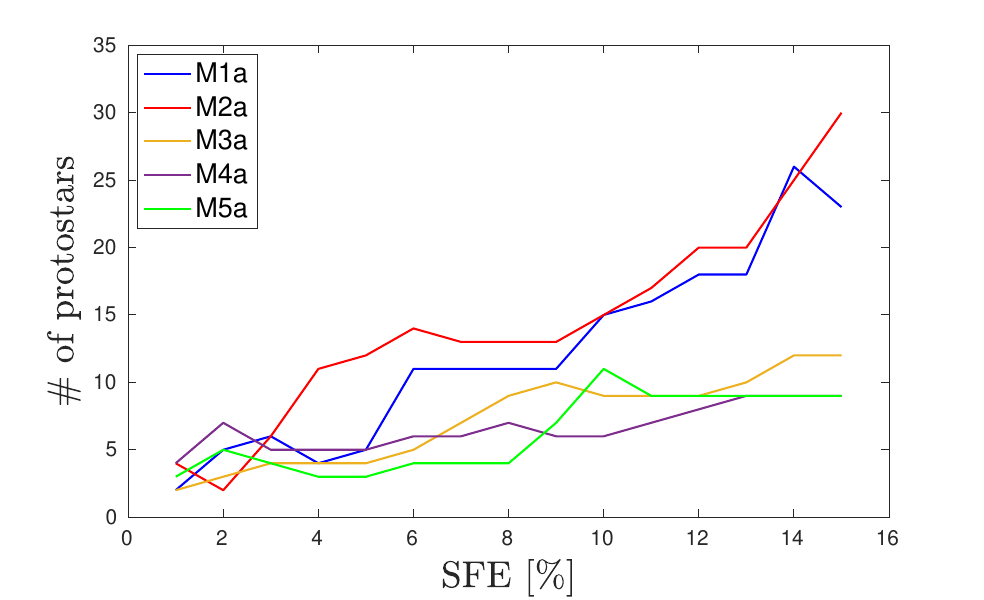}
    \includegraphics[angle=0,scale=0.515]{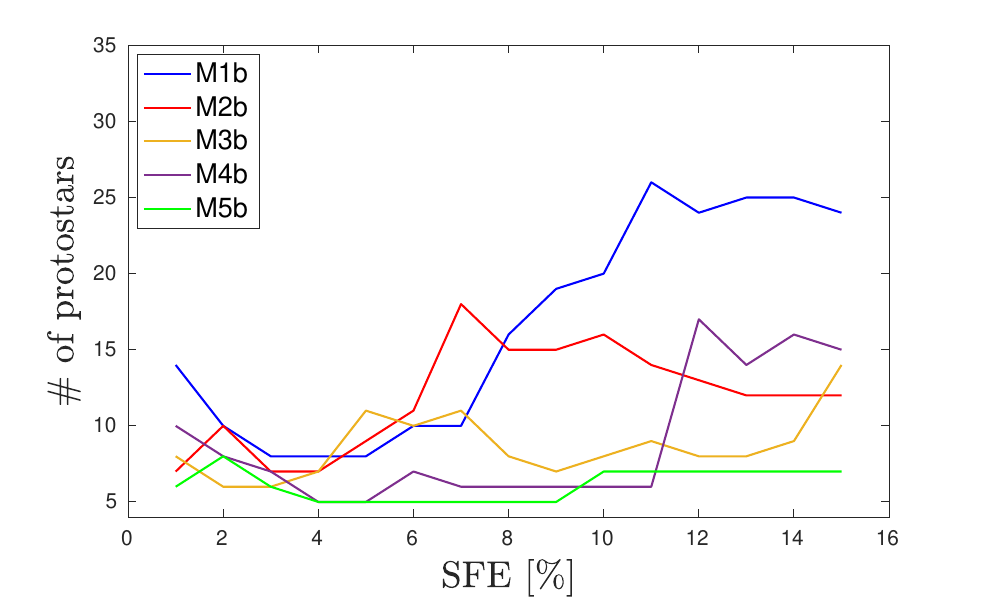}
    \caption{Left - Number of protostars as a function of SFE for the set of models M1a$-$M5a. Right - Number of protostars as a function of SFE for the set of models M1b$-$M5b. Color in online edition. 
    }
  \end{figure*}

            \begin{figure*} \label{fig:9}
	\centering
	\includegraphics[angle=0,scale=0.515]{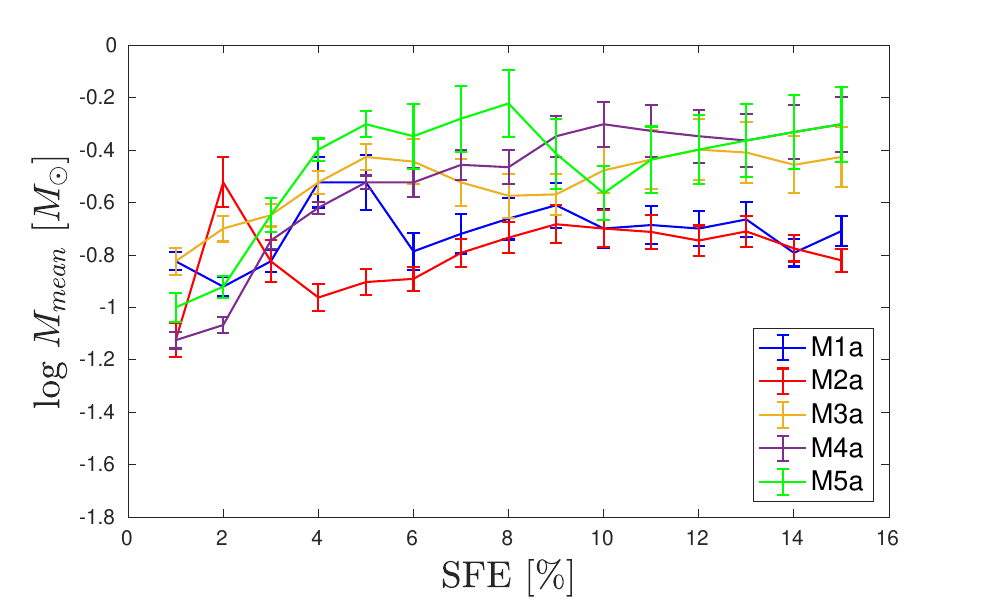}
	\includegraphics[angle=0,scale=0.515]{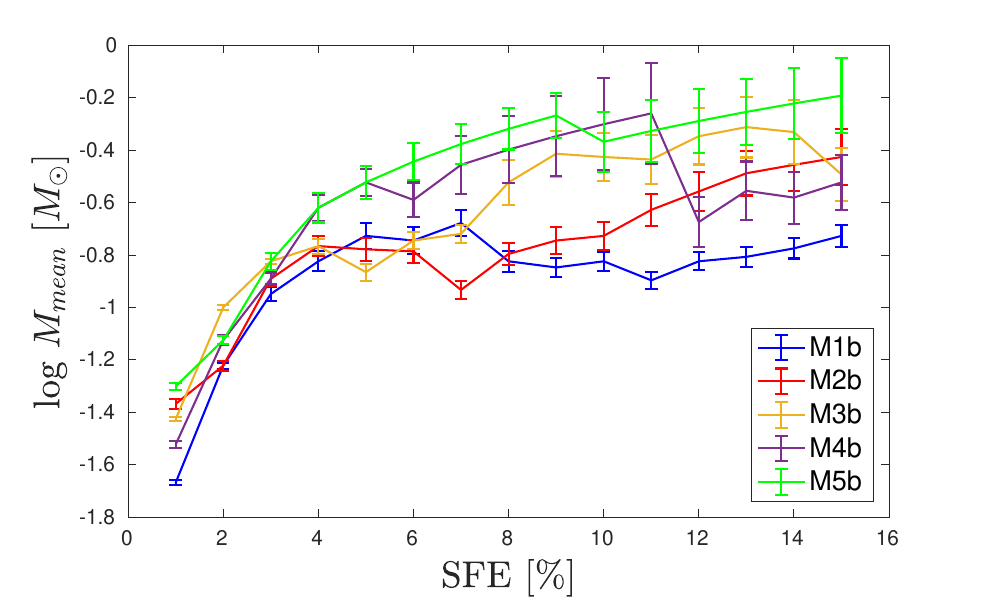}
	\caption{Left - Mean mass of protostars $M^{\ast}_{\rm mean}$ as a function of SFE for set of models M1a$-$M5a. Right - Mean mass of protostars $M^{\ast}_{\rm mean}$ as a function of SFE for set of models M1b$-$M5b. The mean mass of protostars $M^{\ast}_{\rm mean}$ is given in units of solar mass. These profiles represents the stage of evolution where the  SFE in each model reaches $\xi$ = 15 \%. Color in online edition. 
	}
\end{figure*}

          \begin{figure*} \label{fig:7}
    \centering
	\includegraphics[angle=0,scale=0.4]{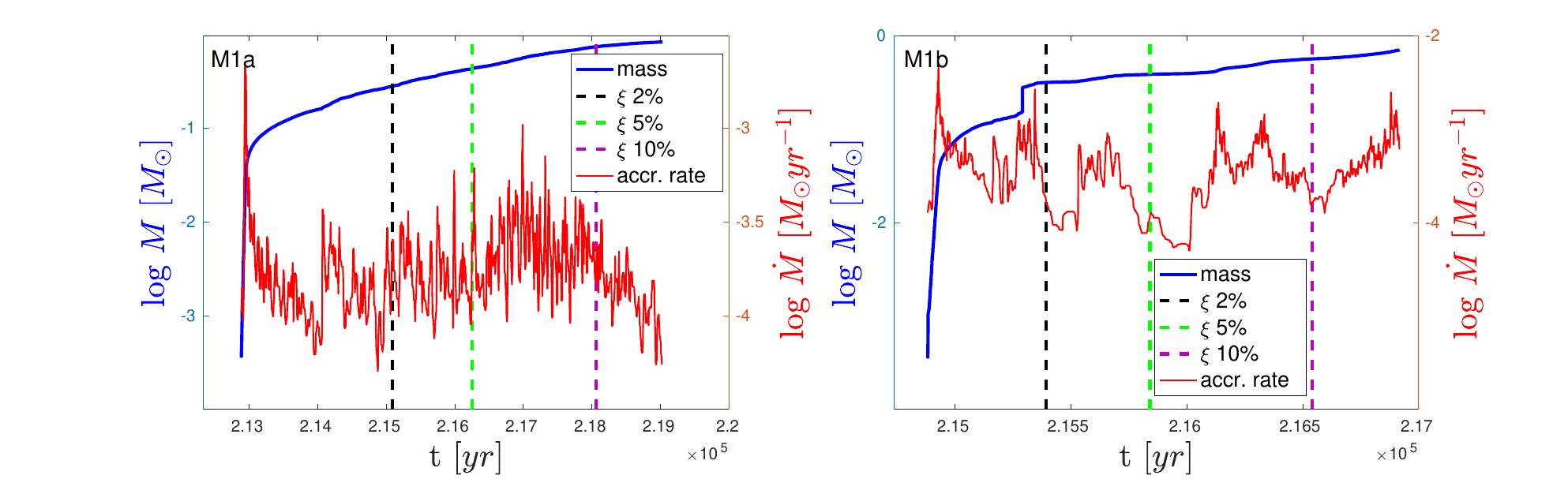}
    \includegraphics[angle=0,scale=0.4]{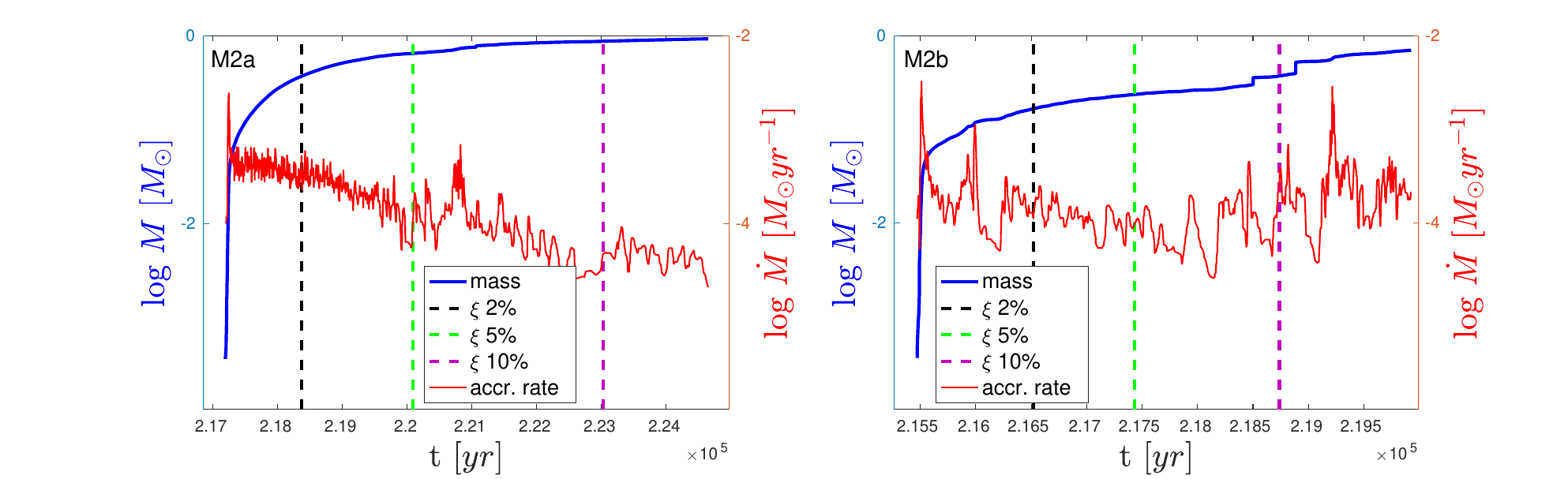}
    \includegraphics[angle=0,scale=0.4]{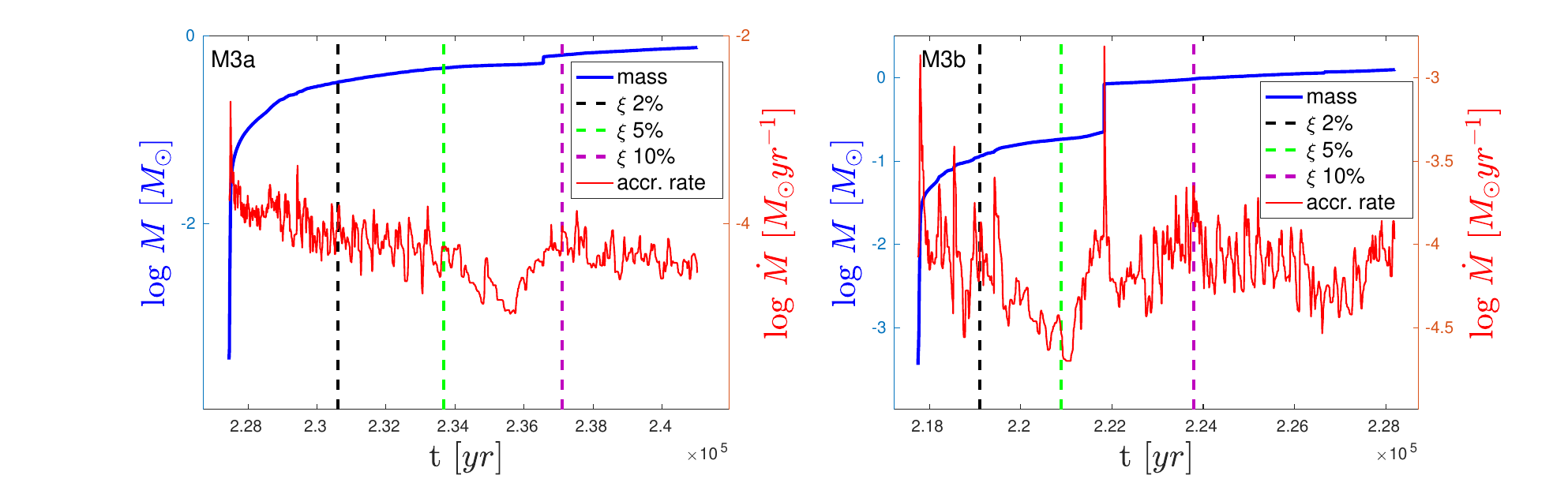}
    \includegraphics[angle=0,scale=0.4]{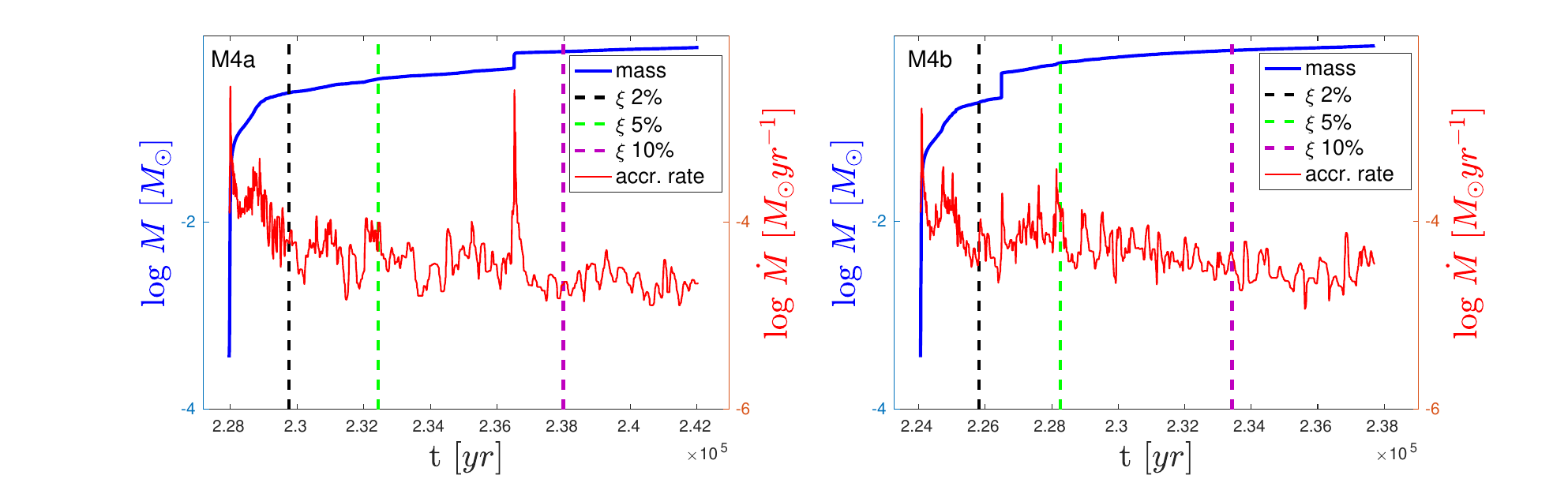}
    \includegraphics[angle=0,scale=0.4]{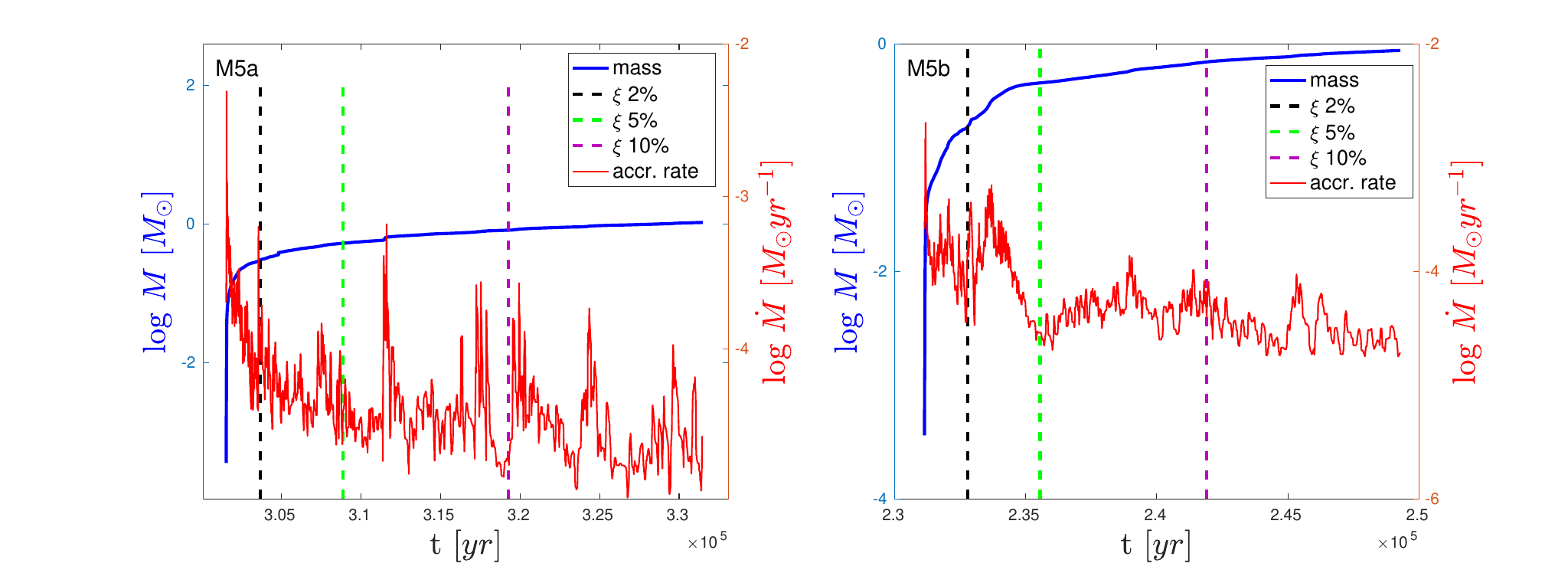}    

    \caption{Mass accumulation history and accretion rate profile for the protostar in each model that survives all merger events and keeps accreting for the longest time. The vertical dashed lines in black, green, purple mark the SFEs $\xi$ = 2 \%, 5 \%, and 10 \%, respectively, whereas the final stage indicates $\xi$ = 15 \% during the model evolution. The accumulated mass is given in solar mass units and the accretion rate in units of solar mass per year while the time is mentioned in years. Color in online edition.
    }
  \end{figure*}


In Figure 4 we show the number of protostars as a function of the SFE. Sometimes the number of protostars increases with the SFE because new protostars form, but we also see dips which are indicative of merger events that keep occurring throughout the gas collapse in each of our models. These mergers impart their deep impact on the statistical evolution of the protostars. The less massive protostars during the dynamical interactions with their counterparts experience a capture phenomenon and are merged with the other protostars, hence, making them more massive and even stronger attractors for the subsequent interactions. This process of dynamical capturing of protostars increases both the mean mass $M^{\ast}_{\rm mean}$ and the value of the most massive protostar $M^{\ast}_{\rm max}$ (see Table 1) residing in the collapsing gas at a given point of time as a function of SFE. Figure 4 (left) shows the evolving total number of emerging protostars as a function of SFE. Figure 4 has two panels in which on the left we show the evolving total number of emerging protostars as a function of SFE  in the respective models M1a$-$M5a. Every sudden rise and fall in the plot indicates the formation of new protostars inside the gas and the mergers of the protostars. We notice that the two processes of emergence and mergers of protostars remain frequent particularly for the cold molecular gas models such as M1a and M2a. From  $\xi$ $\sim$ 5 \% these two cold gas models clearly get separated from the rest of the warm models of M3a$-$M5a which do not show any significant mergers. This behavior continues until $\xi$ $\sim$ 15 \%.     

Similarly, the panel on the right shows the evolving total number of protostars as well as their mergers as a function of SFE in the respective models M1b$-$M5b. Except for the relatively warmer gas cloud model M3b, which shows a marginal lead in terms of the emergence of new protostars and their mergers around ~$\xi$ = 5 \%, the two cold molecular gas models of M1b and M2b again show more protostars compared to the warm gas models M4b and M5b. Model M2b, at least until ~$\xi$ $\sim$ 7 \% remains consistent with the notion that cold gas models yield a greater number of protostars. However, beyond ~$\xi$ $\sim$ 7 \%, the trend is less obvious as statistical fluctuations are significant. From ~$\xi$ $\sim$ 7 \% onwards the trend of a greater number of protostars for cold gas models becomes even more clear, which for the cold gas model of M1b continues until the selected end of the evolving models i.e., $\xi$ = 15 \%.  

The mean mass $M^{\ast}_{\rm mean}$ as a function of SFE in both sets of models is described in Figure 5. The panel on the left illustrates the model set M1a$-$M5a. To understand the results better we first determine the specific value of the SFE where the transition from a fragmentation to accretion dominated regime takes place. On the left, we find $\xi$ $\sim$ 7 \% as the transition point between these two regimes.  

 In the left panel of Figure 5 we find that the mean values of the protostellar mass for the set of models M1a$-$M5a is related to the thermal conditions in the star forming gas (see also Table 1) for $\xi$ $\sim$ 8 \%, providing evidence for higher average masses once accretion takes over in higher temperature models. The total number of fragments is reduced when the initial temperature is higher. It is evident that before $\xi$ $\sim$ 5 \% the mass accumulation in Figure 5 shows no statistically significant dependence on the gas temperature in our results but on dynamical interactions and subsequent mergers. Figure 6 shows the accretion characteristics for the protostar that has the longest accretion history in the collapsing gas cloud without being merged with any other fragments, and in principle a similar behavior should be expected for all protostars on thermal grounds. 
 
 Figure 6 also shows that once the evolution crosses $\xi$ $\sim$ 5$-$8 \%, there is a tendency for enhanced mass accretion in some protostars, although this trend is not shared in all cases. 

In the warm molecular gas clumps are expected to be more massive and also to be further apart because of the fragmentation occurring at lower densities (see Figure 1) and larger separations. Hence in warm gas models mergers are less likely. A significant number of merger events is more likely to happen in cold molecular gas than in warm molecular gas mainly because cold clouds tend to fragment more hence by virtue of dynamical evolution increase the chances of protostellar mergers. After the transition to the accretion dominated regime the mergers continue and may further reduce the number of protostars. This leads to on average more massive protostars in the initially warmer models. Since the accretion dominated phase starts after $\xi$ $\sim$ 5 \%, we see a persistence in the trend of protostars taking higher mean mass values until $\xi$ = 15 \% in all of the explored models. 

Similarly, the right panel in Figure 5 shows the second set of models M1b$-$M5b, which, in general, repeats the trend observed in models M1a$-$M5a. The point of transition from the fragmentation dominated regime to the accretion dominated regime in case of the second set of models appears at $\xi$ $\sim$ 5 \%. In the two sets of models we see that in the $\xi$ = 10 \% to $\xi$ = 15 \% snapshots precisely the warmer clouds yield higher mean masses $M^{\ast}_{\rm mean}$ than clouds with cold molecular gas.

Figure 6 illustrates the mass accumulation history and the mass accretion rate over time for the most long-lived self-gravitating fragment in each of our studied models. To describe the possible trend in the mass accretion rate profiles and their likely connection with the thermal state of the molecular clouds we choose this particular protostar with the intent to show the long-term evolution that takes place during the collapsing phase of star forming gas. The figure shows the mass and the accretion rate of the protostar as a function of time. We also draw three vertical dashed lines in black, green, and purple colors describing the stages of evolution in terms of the SFE of the gas ~at $\xi$ = 2 \%, 5 \%, 10 \%, and 15 \%, respectively. We notice that for a protostar the accretion can be categorized into two types. The first depends on the surrounding temperature of gas that the protostar accretes. For the given gas temperature it follows the relation \citep{Shu1987}

\begin{equation} \label{beta}
\dot M \sim \frac{c_{\rm s}^{3}}{ G },
\end{equation} 
where $c_{\rm s}$ is the sound speed including the dependence on the thermal state of the gas. This is the expected accretion rate in low-mass star formation \citep{Hosokawa2009}. In the second type of accretion the protostar swallows nearby less massive fragment(s) that during the dynamical evolution come closer and start entering the potential well of the protostar. This second type of accretion, which has been referred to as a merger, is usually responsible for the sudden vertical rise in the mass accumulation rate and the associated subsequent peak which occurs in the mass accretion rate of the protostar.

                  \begin{figure*} \label{fig:10}
    \centering
    \includegraphics[angle=0,scale=0.515]{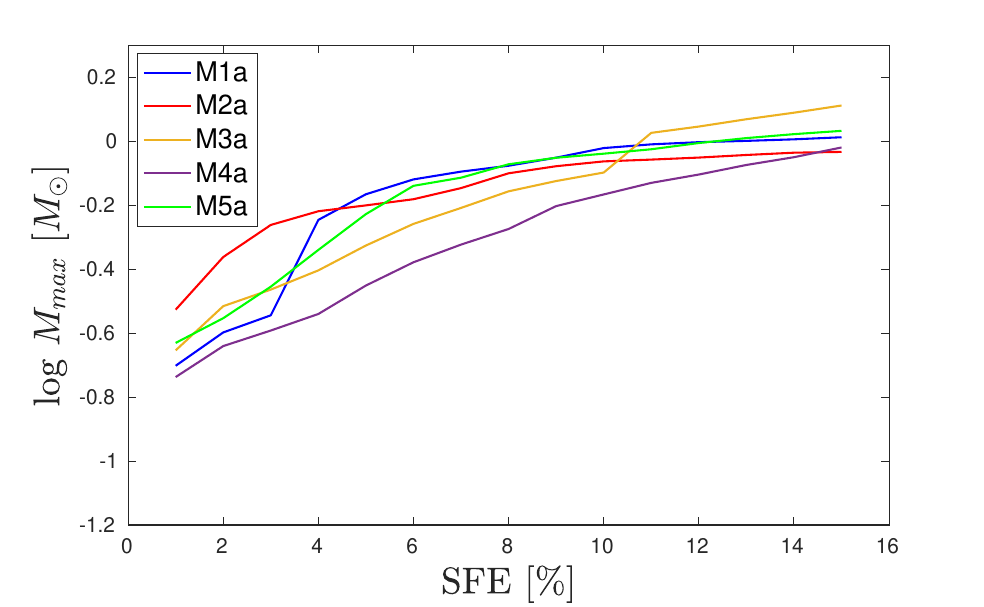}
    \includegraphics[angle=0,scale=0.515]{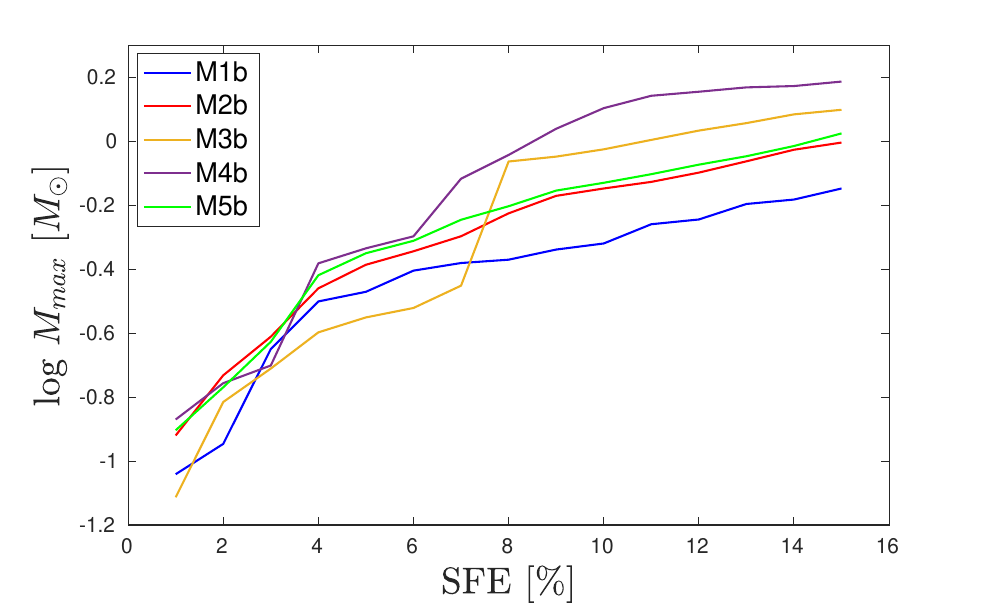}
    \caption{Left - Maximum mass of the protostars as a function of SFE for set of models M1a$-$M5a.
     Right -  Maximum mass of the protostars as a function of SFE for set of models M1b$-$M5b. The maximum masses are given in units of solar mass. Color in online edition. 
    }
  \end{figure*}
  
                    \begin{figure*} \label{fig:11}
    \centering
    \includegraphics[angle=0,scale=0.515]{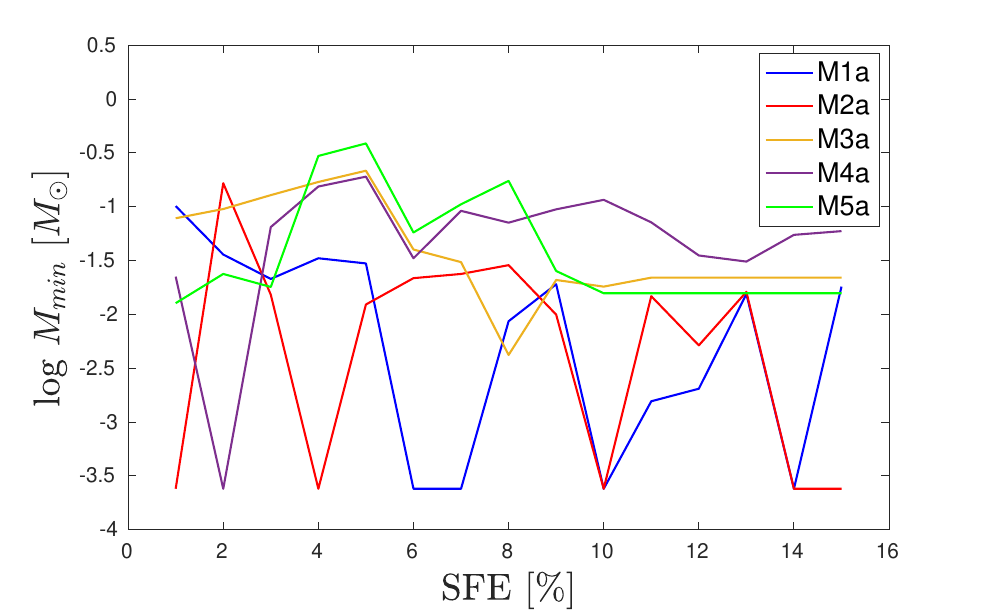}
    \includegraphics[angle=0,scale=0.515]{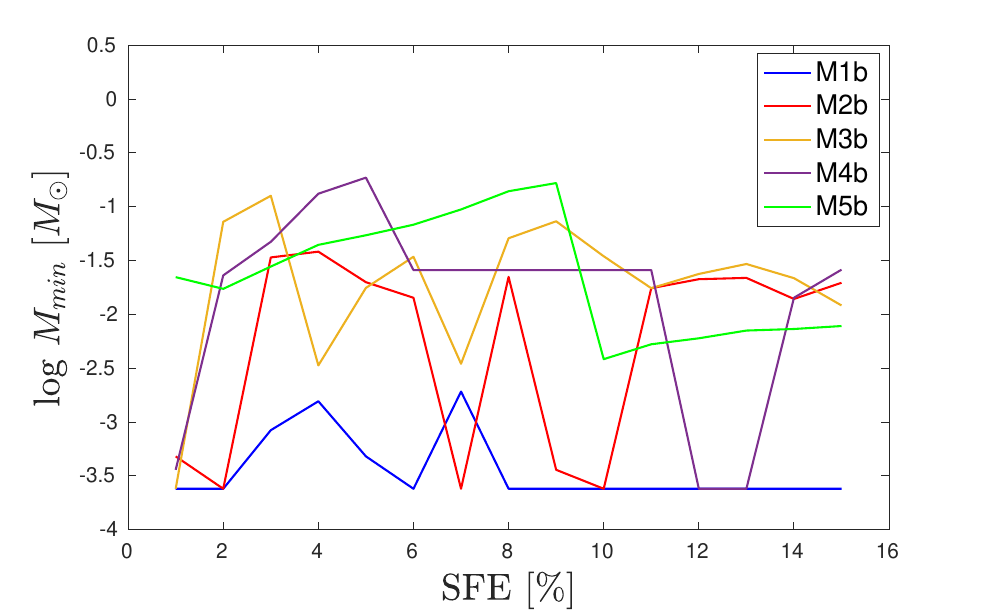}
    \caption{Left - Minimum mass of the protostars as a function of SFE for set of models M1a$-$M5a.
     Right -  Minimum mass of the protostars as a function of SFE for set of models M1b$-$M5b. The minimum masses are given in units of solar mass. Color in online edition.
    }
  \end{figure*}
  

We show the maximum mass $M^{\ast}_{\rm max}$ of the protostars as a function of evolving SFE in each of our models in Figure 7. For the first seed we see in the left panel an initial weak dependence of $M^{\ast}_{\rm max}$ on the initial temperature of the collapsing clouds until  $\xi$ $\sim$ 7 \%. After $\xi$ $\sim$ 7 \% until $\xi$ = 15 \% the protostars in warmer clouds attain higher value of $M^{\ast}_{\rm max}$. Similarly, in the right panel we find an even more clear trend from $\xi$ $\sim$ 5 \% onward with a relatively strong dependence of the maximum stellar mass on the gas temperature. The trend gets pronounced after $\xi$ = 8 \% and continues to be so until $\xi$ = 15 \%.
      
Finally, Figure 8 shows the minimum mass $M^{\ast}_{\rm min}$ of the protostars as a function of evolving SFE in each of our models. This figure is divided into two panels where the left panel indicates models belonging to the first seed and the right panel covers the second seed related models. 
We generally see strong fluctuations in the plots. The most likely reason is that when a protostar is formed, it can only have a mass which is close to the minimum attainable mass in the simulation and so must be of the same order of magnitude as the Jeans mass.  Despite the fluctuations which can temporarily reverse the trend, the lowest protostellar mass tends to increase with higher temperatures in the two panels of the figure.  There is no evidence of a change when the simulations evolve from the fragmentation regime to the accretion dominated regime. 


\section{CONCLUSIONS}

We have good indications that typical features of the IMF such as the mean, minimum and maximum stellar mass are regulated by the two key physical processes of fragmentation and mass accretion. Our simulations indicate the presence of two distinct regimes of protostellar mass growth, one where the protostellar masses are dominated by the initial fragmentation, and one where they are dominated by the accretion process. In the fragmentation dominated regime one expects at best a very weak dependence on the initial temperature of the gas, as the Jeans mass is very similar at the transition point from an approximately isothermal to an adiabatic EOS. In the accretion dominated regime, on the other hand, we find that the average mass correlates with the gas temperature.

We have quantified the role of these processes with numerical simulations, varying the initial gas temperature from 10 to 50 K, assuming transsonic turbulence and a ratio of rotational to gravitational energy of 1 $\%$. We pursued two sets of models with different random seeds to initialize the turbulence, corresponding to different realizations with the same statistical properties. 

Before the transition to the regime dominated by accretion, there is no evidence of a temperature dependence, confirming previous results e.g. by \citet{Lee2018a} and \citet{Lee2019}. As a result, one may expect a rather universal IMF if the SFE is low enough. If higher SFEs are reached, our simulations show that one would expect a dependence of the accretion process on temperature. This could be caused by local radiation backgrounds that heat up the gas. The minimum temperature of the gas is expected to increase with cosmic redshift, as cooling becomes inefficient below the CMB temperature. The temperatures explored here correspond to a redshift range from 2.7 to 17.3, if interpreted to be due to the temperature of the CMB, thus covering a significant range in redshift, while in the presence of a sufficiently strong radiation background heating the gas, the models can be applied at lower redshift as well. Our approach implicitly assumes the presence of dust, as the latter regulates the transition from an approximately isothermal to an adiabatic regime, while a different and more detailed chemical model should be applied for a purely primordial gas \citep[see e.g.][]{Riaz18c}.

The effective mass accretion phase helps the protostars to grow in mass as well as in number which lead to the eventual higher mean masses $M^{\ast}_{\rm mean}$ associated to the warmer clouds until the SFE reaches $\xi$ = 15 \% at the end of our simulations. The total number of protostars in each of our models and the associated protostellar mergers as a function of SFE also provide an insight which support the existence of a transition from a fragmentation dominated to an accretion dominated phase inside collapsing gas clouds. Despite the lesser number of mergers the warmer gas clouds show a higher mean mass $M^{\ast}_{\rm mean}$ after a critical SFE of about $\xi$ $\sim$ 5$-$7 \%. Our analysis of mass accretion for the longest surviving protostar in each model provides a demonstration of the transition from the fragmentation dominated regime to accretion dominated regime in star forming gas clouds. 

The maximum mass of the protostars  also seems to follow the transition from the fragmentation to the accretion dominated phase. $M^{\ast}_{\rm max}$ shows a relatively strong dependence on the temperature of the cloud and also reflects the critical SFE within $\xi$ $\sim$ 5$-$7 \%. The minimum mass of the protostars in our explored models, however, shows a much weaker dependence and remains subject to statistical fluctuations. We suspect that such a trend, for the maximum mass in particular, could be an indication that the increasing thermal pressure in the star forming gas removes stars from the low-mass end of the IMF. Overall, our results suggest that the IMF will be influenced by the initial temperature of the gas if the SFR is high enough.

\section{Acknowledgements}
This research was partially supported by the supercomputing infrastructure of the NLHPC (ECM$-$02). The first author RR also acknowledges the high performance computing cluster Kultrun. RR and the second author DRGS thank for funding through Fondecyt Postdoctorado (project code 3190344) and the Concurso Proyectos Internacionales de Investigaci\'on, Convocatoria 2015" (project code PII20150171). DRGS further thanks for funding via Fondecyt regular (project code 1161247) and via the Chilean BASAL Centro de Excelencia en Astrof\'isica yTecnolog\'ias Afines (CATA) grant PFB-06/2007. This work was also funded by the CONICYT PIA ACT172033.  SV wishes to thank Prof. Dr. R. Keppens and Prof.  Dr. S. Poedts for providing access to the KUL supercomputing cluster Thinking while developing and testing the code that was used in this work. He also gratefully acknowledges the support of the KUL HPC team. RSK acknowledges financial support from the German Research Foundation (DFG) via the collaborative research centre (SFB 881, Project-ID 138713538) ``The Milky Way System'' (subprojects B1, B2, and B8) and from the Heidelberg cluster of excellence EXC 2181(Project-ID 390900948) ``STRUCTURES: A unifying approach to emergent phenomena in the physical world, mathematics, and complex data'' funded by the German Excellence Strategy.         


\label{lastpage}

\end{document}